\documentclass{article}\usepackage[]{graphicx}\usepackage[]{color}
\makeatletter
\def\maxwidth{ %
  \ifdim\Gin@nat@width>\linewidth
    \linewidth
  \else
    \Gin@nat@width
  \fi
}
\makeatother

\definecolor{fgcolor}{rgb}{0.345, 0.345, 0.345}
\makeatletter
\@ifundefined{AddToHook}{}{\AddToHook{package/xcolor/after}{\definecolor{fgcolor}{rgb}{0.345, 0.345, 0.345}}}
\makeatother

\usepackage{framed}
\makeatletter
\newenvironment{kframe}{%
 \def\at@end@of@kframe{}%
 \ifinner\ifhmode%
  \def\at@end@of@kframe{\end{minipage}}%
  \begin{minipage}{\columnwidth}%
 \fi\fi%
 \def\FrameCommand##1{\hskip\@totalleftmargin \hskip-\fboxsep
 \colorbox{shadecolor}{##1}\hskip-\fboxsep
     \hskip-\linewidth \hskip-\@totalleftmargin \hskip\columnwidth}%
 \MakeFramed {\advance\hsize-\width
   \@totalleftmargin\z@ \linewidth\hsize
   \@setminipage}}%
 {\par\unskip\endMakeFramed%
 \at@end@of@kframe}
\makeatother

\definecolor{shadecolor}{rgb}{.97, .97, .97}
\definecolor{messagecolor}{rgb}{0, 0, 0}
\definecolor{warningcolor}{rgb}{1, 0, 1}
\definecolor{errorcolor}{rgb}{1, 0, 0}
\makeatletter
\@ifundefined{AddToHook}{}{\AddToHook{package/xcolor/after}{
\definecolor{shadecolor}{rgb}{.97, .97, .97}
\definecolor{messagecolor}{rgb}{0, 0, 0}
\definecolor{warningcolor}{rgb}{1, 0, 1}
\definecolor{errorcolor}{rgb}{1, 0, 0}
}}
\makeatother
\newenvironment{knitrout}{}{} 

\usepackage{alltt}
\usepackage{amsmath,amsfonts,amssymb}
\usepackage{graphicx}
\usepackage{natbib}
\usepackage{hyperref}
\newtheorem{proposition}{Proposition}
\newtheorem{corollary}{Corollary}

\newcommand{\Prob}[1]{\mathbb{P}\left( #1 \right) }
\newcommand{\PriorMeasure}{\Pi_0}
\newcommand{\priordensity}{\pi_0}
\newcommand{\PosteriorMeasure}{\Pi_1}
\newcommand{\posteriordensity}{\pi_1}
\newcommand{\WeightedApproxPost}{\widehat{\Pi}_1}
\newcommand{\Var}[1]{\mathrm{Var}\left[ #1 \right]}
\newcommand{\Cov}[1]{\mathrm{Cov}\left[ #1 \right]}

\newcommand{\ReplicatedApproxPost}{\tilde{\Pi}_1}
\newcommand{\SampledApproxPost}{\Pi^{\dagger}_1}
\newcommand{\Renyi}{R{\'e}nyi}
\newcommand{\myexp}[1]{\exp{\left\{ #1 \right\}}}
\newcommand{\MLE}{\hat{\theta}}

\title{Evaluating Posterior Distributions by Selectively Breeding Prior Samples}
\author{Cosma Rohilla Shalizi\thanks{Departments of Statistics and Data Science, and of Machine Learning, Carnegie Mellon University, Pittsburgh, PA 15213 USA, and the Santa Fe Institute, 1399 Hyde Park Road, Santa Fe, NM 87501}}
\date{24 February 2016, last \LaTeX 'd \today}
\IfFileExists{upquote.sty}{\usepackage{upquote}}{}
\begin{document}
\maketitle

\begin{abstract}
  Using Markov chain Monte Carlo to sample from posterior
  distributions was the key innovation which made Bayesian data
  analysis practical.  Notoriously, however, MCMC is hard to tune,
  hard to diagnose, and hard to parallelize.  This pedagogical note
  explores variants on a universal {\em non}-Markov-chain Monte Carlo
  scheme for sampling from posterior distributions.  The basic idea is
  to draw parameter values from the prior distributions, evaluate the
  likelihood of each draw, and then copy that draw a number of times
  proportional to its likelihood.  The distribution after copying is
  an approximation to the posterior which becomes exact as the number
  of initial samples goes to infinity; the convergence of the
  approximation is easily analyzed, and is uniform over
  Glivenko-Cantelli classes.  While not {\em entirely} practical, the
  schemes are straightforward to implement (a few lines of R), easily
  parallelized, and require no rejection, burn-in, convergence
  diagnostics, or tuning of any control settings.  I provide
  references to the prior art which deals with some of the practical
  obstacles, at some cost in computational and analytical simplicity.
\end{abstract}

Suppose we have a prior probability measure $\PriorMeasure$ on $\Theta$, and a
likelihood function $f(\theta,x)$.  (Since $x$ plays little further role in
what follows, I suppress it in the notation.)  We want to sample from the
posterior measure $\Pi_1$, defined via Bayes's rule:
\[
\PosteriorMeasure(A) = \frac{\PriorMeasure(fA)}{\PriorMeasure(f)}
\]
using the de Finetti notation for expectations, so that $\PriorMeasure(fA) =
\int{f(\theta)\mathbf{1}_{A}(\theta) d\PriorMeasure(\theta)}$.  I explain how
to obtain a sample from $\PosteriorMeasure$, supposing only that we can draw an
IID sample from $\PriorMeasure$, and that we can calculate the likelihood.  I
first develop a weighted sampling scheme, and then consider ways of producing
unweighted samples.

The inspiration for the algorithms below lies in the following observation.  If
$\PriorMeasure$ and $\PosteriorMeasure$ have densities $\priordensity$ and
$\posteriordensity$, then Bayes's rule takes the form
\[
\posteriordensity(\theta) = \frac{f(\theta)\priordensity(\theta)}{\PriorMeasure(f)}
\]
and this is exactly the form of the ``replicator equation'' of evolutionary
biology, which describes the effects of natural selection on the gene pool of a
population \citep[Appendix A]{CRS-dynamics-of-bayes}.  In the evolutionary
interpretation, $\theta$ refers to genotypes rather than parameter values, and
$f(\theta)$ is not likelihood but ``fitness'', the mean number of surviving
offspring left by individuals of genotype $\theta$
\citep{Hofbauer-Sigmund-evol-games-and-pop-dyn}.  Usually, in biology, the
fitness of any one genotype is ``frequency-dependent'', i.e., a function of the
whole distribution of genotypes, but this is excluded in Bayes's rule.

Bayesian updating describes an evolutionary process in an infinitely large
population without competition, cooperation, mutation or sex.  The algorithms
below are all variants on taking this literally with a finite population.

Very similar algorithms have of course been published before; \S
\ref{sec:related} provides references, and \S\S \ref{sec:pros}--\ref{sec:cons}
discusses some of their pros and cons.

\section{Fitness-Weighted Samples}

The fitness-weighted algorithm is very simple.

\paragraph{Algorithm 1: Likelihood Importance Prior Sampling (LIPS)}

\begin{enumerate}
\item Draw $n$ samples $\theta_1, \theta_2, \ldots \theta_n$ iidly from $\PriorMeasure$.
\item For each $\theta_i$, calculate the likelihood $f(\theta_i)$.
\item Approximate $\PosteriorMeasure$ by
\begin{equation}
\WeightedApproxPost = \frac{\sum_{i=1}^{n}{f(\theta_i) \delta_{\theta_i}}}{\sum_{j=1}^{n}{f(\theta_j)}} \label{eqn:weighted-approximate-posterior-measure}
\end{equation}
\end{enumerate}

In particular, the posterior probability of any measurable set $A \subset
\Theta$ is approximated by
\begin{equation}
\WeightedApproxPost(A) = \frac{\sum_{i=1}^{n}{f(\theta_i) \mathbf{1}_{A}(\theta_i)}}{\sum_{j=1}^{n}{f(\theta_j)}} \label{eqn:weighted-approximate-posterior-prob}
\end{equation}

Clearly, $\WeightedApproxPost$ is a probability measure.  It converges on
$\PosteriorMeasure$ as $n$ grows:

\begin{proposition}
  Suppose $\PosteriorMeasure(f) > 0$.  Then as $n \rightarrow \infty$,
  $\WeightedApproxPost(A) \rightarrow \PosteriorMeasure(A)$ almost surely
  $(\PriorMeasure)$.
  \label{prop:basic-almost-sure-convergence}
\end{proposition}

\textsc{Proof:} Divide both the numerator and denominator of Eq.\
\ref{eqn:weighted-approximate-posterior-prob} by $n$:
\[
\WeightedApproxPost(A) = \frac{n^{-1} \sum_{i=1}^{n}{f(\theta_i)
    \mathbf{1}_{A}(\theta_i)}}{n^{-1}\sum_{j=1}^{n}{f(\theta_j)}}
\]
Call the numerator $N_A$ and the denominator $D$.  Since $\theta_i$ are IID, by
the law of large numbers, $N_A \rightarrow \PriorMeasure(fA)$ and $D
\rightarrow \PriorMeasure(f)$ almost surely.  By the continuous mapping theorem
for almost-sure convergence, then, $N_A/D \rightarrow
\PriorMeasure(fA)/\PriorMeasure(f) = \PosteriorMeasure(A)$.  $\Box$

By the usual measure-theoretic arguments, Proposition
\ref{prop:basic-almost-sure-convergence} extends from measurable sets to
integrable functions.

\begin{corollary}
  Suppose $\PosteriorMeasure(f) > 0$, and that $\Theta$ is a locally compact,
  second-countable Hausdorff (lcscH) space, such as $\mathbb{R}^d$.  Then as
  $n\rightarrow\infty$, $\WeightedApproxPost \stackrel{D}{\rightarrow}
  \PosteriorMeasure$ almost surely $(\PriorMeasure)$.
\end{corollary}

\textsc{Proof:} Take any finite collection of sets $A_1, A_2, \ldots A_k$ which
are all continuity sets for $\PosteriorMeasure$, i.e., sets whose boundaries
$\partial A_i$ have $\PosteriorMeasure$ measure 0.  By Proposition
\ref{prop:basic-almost-sure-convergence}, $\WeightedApproxPost(A_i) \rightarrow
\PosteriorMeasure(A_i)$ a.s. ($\PriorMeasure$).  For any finite $k$, therefore,
$(\WeightedApproxPost(A_1), \ldots \WeightedApproxPost(A_k)) \rightarrow
(\PosteriorMeasure(A_1), \ldots \PosteriorMeasure(A_k))$
a.s. ($\PriorMeasure$).  Since the convergence holds almost surely, {\em a
  fortiori} it holds in distribution.  But now by \citet[Theorem
16.16(iii)]{Kallenberg-mod-prob}, $\WeightedApproxPost
\stackrel{D}{\rightarrow}\PosteriorMeasure$.  $\Box$

There is also a Glivenko-Cantelli-type result.

\begin{corollary}
  Suppose that $\Theta$ is Borel-isomorphic to a complete, separable metric
  (``Polish'') space, and that $\mathcal{F}$ is a bounded, separable family of
  measurable functions on $\Theta$, where either the bracketing or covering
  number\footnote{See, e.g., \citet{Anthony-Bartlett-neural-network-learning}
    or \citet{van-de-Geer-empirical} for the definitions of bracketing and
    covering numbers.}  of $\mathcal{F}$ is finite under every measure on
  $\Theta$.  Then, with $\PriorMeasure$-probability 1,
  \[
  \sup_{f \in \mathcal{F}}{\left| \WeightedApproxPost(f) -
      \PosteriorMeasure(f)\right|} \rightarrow 0
  \]
\end{corollary}

\textsc{Proof:} Finite bracketing number and finite covering number are
equivalent to each other and to the universal Glivenko-Cantelli property, by
\citet[Theorem 1.3]{van-Handel-universal-GC}.  The combination of these
assumptions with Proposition \ref{prop:basic-almost-sure-convergence} thus
meets sufficient condition (6) of \citet[Corollary
1.4]{van-Handel-universal-GC}, and the result follows.  $\Box$

\subsection{Large-Sample Asymptotics}
\label{sec:asymptotics}

We may say more about the rate of convergence of $\WeightedApproxPost(A)$ to
$\PosteriorMeasure(A)$.

\begin{proposition}
Suppose $\PriorMeasure(f) > 0$.  Then
\[
n\Var{\WeightedApproxPost(A)} \rightarrow \frac{\PriorMeasure(f^2)\PosteriorMeasure^2(A) + \PriorMeasure(f^2A)(1-2\PosteriorMeasure(A))}{\PriorMeasure^2(f)}
\]
where the variance is taken over the drawing of the $\theta_i$ from
$\PriorMeasure$.
\end{proposition}

\textsc{Proof:} Defining $N_A$ and $D$ as in the proof of Proposition
\ref{prop:basic-almost-sure-convergence}, the delta method gives an asymptotic
variance:
\begin{equation}
\Var{\WeightedApproxPost(A)} \rightarrow {\left(\frac{\partial \WeightedApproxPost(A)}{\partial N_A}\right)}^2 \Var{N_A} + {\left(\frac{\partial \WeightedApproxPost(A)}{\partial D}\right)}^2 \Var{D} + 2 \frac{\partial \WeightedApproxPost(A)}{\partial N_A} \frac{\partial \WeightedApproxPost(A)}{\partial D} \Cov{N_A,D}
\end{equation}

Fill in each piece, using the fact that an indicator function is equal to its own square:
\begin{eqnarray}
\frac{\partial \WeightedApproxPost(A)}{\partial N_A} & = & \frac{1}{D} \rightarrow \frac{1}{\PriorMeasure(f)}\\
\frac{\partial \WeightedApproxPost(A)}{\partial D} & = & -\frac{N_A}{D^2} \rightarrow -\frac{\PriorMeasure(fA)}{\PriorMeasure^2(f)}\\
\Var{N_A} & = & n^{-1}(\PriorMeasure(f^2A^2) - \PriorMeasure^2(fA)) = n^{-1} (\PriorMeasure(f^2A) - \PriorMeasure^2(fA))\\
\Var{D} & = & n^{-1}(\PriorMeasure(f^2) - \PriorMeasure^2(f))\\
\Cov{N_A, D} & = & n^{-1}(\PriorMeasure(f^2A) - \PriorMeasure(fA)\PriorMeasure(f))
\end{eqnarray}
Put these together, and pull out the common factor of $n$:
\begin{eqnarray}
n\Var{\WeightedApproxPost(A)} & \rightarrow & \frac{\PriorMeasure(f^2A) - \PriorMeasure^2(fA)}{\PriorMeasure^2(f)}\\
\nonumber & & + \frac{\PriorMeasure^2(fA) (\PriorMeasure(f^2) - \PriorMeasure^2(f))}{\PriorMeasure^4(f)}\\
\nonumber & & - 2 \frac{\PriorMeasure(fA)}{\PriorMeasure^3(f)}(\PriorMeasure(f^2A) - \PriorMeasure(fA)\PriorMeasure(f))
\end{eqnarray}
Since $\PosteriorMeasure(A) = \PriorMeasure(fA)/\PriorMeasure(f)$, after some algebra,
\[
n\Var{\WeightedApproxPost(A)} \rightarrow  \PosteriorMeasure^2(A)\frac{\PriorMeasure(f^2)}{\PriorMeasure^2(f)} + \frac{\PriorMeasure(f^2 A)(1-2\PosteriorMeasure(A))}{\PriorMeasure^2(f)}
\]
as was claimed. $\Box$

Remarks, or sanity checks: (i) $1 \geq 1-2\PosteriorMeasure(A) \geq -1$, and
$\PriorMeasure(f^2) \geq \PriorMeasure(f^2A)$, so the numerator is always $\geq
0$.  (ii) As $A \rightarrow \emptyset$ or $A \rightarrow \Theta$, the variance
$\rightarrow 0$.  (iii) As $\PosteriorMeasure(A) \rightarrow 1$, the variance
tends to $\PriorMeasure(f^2 A^c)/\PriorMeasure^2(f)$, while as
$\PosteriorMeasure(A) \rightarrow 0$, the numerator approaches
$\PriorMeasure(f^2 A)/\PriorMeasure^2(f)$.  (iv) An oracle which could sample
directly from $\PosteriorMeasure$ would have an asymptotic variance
proportional to $\PosteriorMeasure(A)-\PosteriorMeasure^2(A)$, so the
algorithm is not as efficient as the oracle.

\begin{corollary}
Suppose $\PriorMeasure(f) > 0$.  Then for any set $A$,
\[
\lim_{n}{n\Var{\WeightedApproxPost(A)}}  \leq 2\frac{\PriorMeasure(f^2)}{\PriorMeasure^2(f)} = 2\exp{D_2(\PosteriorMeasure\|\PriorMeasure)}
\]
where $D_2$ is the \citet{Renyi-introduces-Renyi-entropy} divergence of order 2.
\label{corollary:asymptotic-variance-and-renyi-divergence}
\end{corollary}

\textsc{Proof:} Since $ \PriorMeasure(f^2 A)(1-2\PosteriorMeasure(A)) \leq
\PriorMeasure(f^2)$, the previous proposition gives us an asymptotic variance
of at most $2\PriorMeasure(f^2)/\PriorMeasure^2(f)$ for any set.  Now, the
\Renyi\ divergence of order $\alpha \geq 0$ is defined (for $\alpha \neq 0, 1,
\infty$) to be \citep{van-Erven-Harremoes-renyi-divergence}
\[
D_{\alpha}(P\|Q) = \frac{1}{\alpha-1}\log{P\left({\left(\frac{dP}{dQ}\right)}^{\alpha-1}\right)}
\]
Pick $\alpha=2$, and suppose without loss of generality that $P$ and $Q$ have
densities $p$ and $q$ with respect to a common measure $M$, such as
$(P+Q)/2$, so $\frac{dP}{dQ}(\theta) = p(\theta)/q(\theta)$.  Then
\[
D_2(P\|Q) = \log{P\left(\frac{p}{q}\right)} = \log{M\left( \frac{p^2}{q}\right)} = \log{Q\left(\frac{p^2}{q^2}\right)}
\]
Set $P=\PosteriorMeasure$ and $Q=\PriorMeasure$.  The density ratio is
\[
\frac{p(\theta)}{q(\theta)} = \frac{f(\theta)}{\PriorMeasure(f)}
\]
so
\[
D_2(\PosteriorMeasure\|\PriorMeasure) = \log{\PriorMeasure\left(\frac{f^2(\theta)}{\PriorMeasure^2(f)}\right)} = \log{\frac{\PriorMeasure(f^2)}{\PriorMeasure^2(f)}}
\]
Undoing the logarithm proves the result. $\Box$

\subsection{High-Information Asymptotics}
\label{sec:laplace}

The preceding results deal with the behavior as the number of Monte Carlo
samples $n$ grows; to sum up (Corollary
\ref{corollary:asymptotic-variance-and-renyi-divergence}), with a large number
of draws from the prior, the accuracy of the approximation is controlled by
$\PriorMeasure(f^2)/\PriorMeasure^2(f) = 1 + \Var{f}/\PriorMeasure^2(f)$.
There is also a second asymptotic regime of interest, which is when the
observation $x$ becomes highly informative.  It would be somewhat perverse if
the approximation broke down just when the observation became most useful.

These fears can be calmed, at least in some situations, by asymptotic
expansions.  Suppose that $\Theta=\mathbb{R}^d$ and $\PriorMeasure$ has a
density $\priordensity$ with respect to Lebesgue measure, so $\PriorMeasure(g)
= \int{g(\theta) \priordensity(\theta)d\theta}$.  In particular, write
\[
\PriorMeasure(f) = \int{\priordensity(\theta)\myexp{-t\ell(\theta)} d\theta}
\]
and
\[
\PriorMeasure(f^2) = \int{\priordensity(\theta)\myexp{-2t\ell(\theta)} d\theta}
\]
where $t$ gauges, in some relevant sense, how informative the observation $x$
is, with the normalized negative log-likelihood $\ell(\theta)$ converging to a
constant-magnitude function as $t\rightarrow\infty$.  ($t$ may represent the
number of measurements taken, the length of a time series, the precision of a
measuring instrument, etc.).  Assume that $\ell(\theta)$ has a unique interior
minimum at the MLE $\MLE$.  We may then appeal to Laplace approximation to
evaluate these integrals: by Eq.\ 2.3 in \citet{Tierney-Kass-Kadane}:
\[
\PriorMeasure(f) = (2\pi)^{d/2} |\Sigma|^{1/2} f(\MLE) \left(\priordensity(\MLE) + A_1t^{-1} \right) + O(t^{-2})
\]
where $\Sigma$ is the inverse Hessian matrix of $\ell$ at $\MLE$.  The
precise expression for $A_1$ in terms of the derivatives of $\posteriordensity$
and $\ell$ is known
\citep{Wojdylo-on-Laplace-approximation,Nemes-on-Laplace-approximation} but
not, for the present purposes, illuminating.  Appealing to Laplace
approximation a second time,
\[
\PriorMeasure(f^2) = (2\pi)^{d/2}|\Sigma/2|^{1/2}f^2(\MLE) \left(\priordensity(\MLE) + A_2t^{-1} \right) + O(t^{-2})
\]
Thus
\begin{eqnarray}
  \frac{\PriorMeasure(f^2)}{\PriorMeasure^2(f)} & = & \frac{(2\pi)^{d/2}|\Sigma/2|^{1/2}f^2(\MLE) \left(\priordensity(\MLE) + A_2t^{-1} \right) + O(t^{-2})}{\left[(2\pi)^{d/2} |\Sigma|^{1/2} f(\MLE) \left(\priordensity(\MLE) + A_1t^{-1} \right) + O(t^{-2}) \right]^2}\\
  & = & (4\pi)^{-(d/2)} \frac{\priordensity(\MLE) + A_1 t^{-1} + O(t^{-2})}{\priordensity^2(\MLE) + 2A_2 \priordensity(\MLE) t^{-1} + O(t^{-2})}\\
  & = & \frac{(4\pi)^{-(d/2)}}{\priordensity(\MLE)}\frac{1+ \frac{A_1}{\priordensity(\MLE)}t^{-1} + O(t^{-2})}{1 + 2 \frac{A_2}{\priordensity(\MLE)}t^{-1} + O(t^{-2})}\\
  & = & \frac{(4\pi)^{-(d/2)}}{\priordensity(\MLE)}\left(1 + \frac{A_1 - 2 A_2}{\posteriordensity(\MLE)t}\right) + O(t^{-2})
\end{eqnarray}
The upshot of this is calculation\footnote{Marginally further simplified, when
  $d=1$, by the fact that then $A_2 = A_1/2$; whether this is true more
  generally I do not know.} is that as $t\rightarrow\infty$, that is, as the
data becomes increasingly informative, the ratio
$\PriorMeasure(f^2)/\PriorMeasure^2(f)$ tends to a limit proportional to the
prior density at the MLE, with vanishing corrections.  Heuristically, as
$t\rightarrow\infty$, both $\PriorMeasure(f^2)$ and $\PriorMeasure(f)$ become
integrals dominated by their slowest-decaying integrand.  The rate of
exponential decay is twice as fast for $f^2$ as for $f$, so
$\PriorMeasure(f^2)/\PriorMeasure^2(f)$ tends to a $t$-independent
limit.\footnote{\citet{Bengtsson-Bickel-Li-on-particle-filters} analyzes a
  related problem, in the context of particle filtering, and concludes that
  increasing $t$, as from a growing number of observations, leads to
  ``collapse'' of the filter, in the sense that
  $\PriorMeasure(\max_{i}{f(\theta_i)/D}) \rightarrow 1$, unless $n$ grows
  exponentially in some (fractional) power of $t$.  They do not, however,
  analyze the quality of the approximation {\em to the posterior distribution},
  which is my concern here.  The limit they consider is one where the
  likelihood is extremely sharply peaked over a region of very small prior
  probability mass, so it is indeed unsurprising that if the number of samples
  is small, only a few fall within it, and many samples would be needed to
  accurately sketch a very rapidly-changing posterior.}  The real source of a
difficulty, if there is one, is $\priordensity(\MLE)$ being small.

It's worth remarking that (in the present notation)
$\log{\posteriordensity(\MLE)/\priordensity(\MLE)}$ was proposed by
\citet[pp.\ 483, 486]{Haldane-measurement-of-natural-selection} as a measure of the intensity
of natural selection on a population.

\subsection{Sketch of Finite-Sample Bounds}

It would be nice if there were finite-sample concentration results to go along
with these asymptotics.  One approach to this would be to use results developed
for weighting training samples to cope with distributional shift.  The most
nearly applicable result is that of \citet[Theorem 7 and Corollary
1]{Cortes-Mansour-Mohri-bounds}, which runs (in the present notation) as
follows.  If the samples $\theta_1, \ldots \theta_n$ are drawn iidly from $Q$,
and $dP/dQ = w$, then over any function class $H$ bounded between 0 and 1,
\[
\Prob{\sup_{h\in H}{\frac{P(h) -
      n^{-1}\sum_{i=1}^{n}{w(\theta_i)
        h(\theta_i)}}{\sqrt{Q(w^2)}}} > \epsilon} \leq 4 G_H(2n)
e^{-n\epsilon^{8/3}/4^{5/3}}
\]
where $G_H$ is the growth function of $H$, the number of ways a set of $n$
points can be dichotomized by sets from $H$
\citep{Anthony-Bartlett-neural-network-learning}.  As in this result, we sample
from one distribution and want to compute integrals under another probability
measure, so we would like to set $Q=\PriorMeasure$ and $P=\PosteriorMeasure$.
(This would make the denominator on the left hand side
$\myexp{D_2(\PosteriorMeasure\|\PriorMeasure)/2}$.) Standing in the way is the
annoyance that the conversion weights $w =f(\theta)/\PriorMeasure(f)$ are
unavailable, because instead of $\PriorMeasure(f)$ we only have a Monte Carlo
approximation $D$.  Introduce a new random measure $R$, defined through
$dR/d\PriorMeasure = f/D$.  ($R$ is not, generally, a properly normalized
probability measure.)  The theorem of \citeauthor{Cortes-Mansour-Mohri-bounds}
then bounds the deviation of $\WeightedApproxPost(h)$ from $R(h)$.  Since
$d\PosteriorMeasure/dR = D/\PriorMeasure(f)$, a finite-sample deviation
inequality for $D$ will yield a bound on the relative deviation of
$\WeightedApproxPost(h)$ from $\PosteriorMeasure(h)$.  Controlling the
convergence of $D$ to $\PriorMeasure(f)$ will need assumptions on the
distribution (under the prior) of likelihoods, which seem too model-specific to
pursue here.

\section{Unweighted Samples}

Weighted samples are fine for many purposes, and simple to analyze, but there
are times when it is nicer to not have to keep track of weights.  There are
several ways of modifying the algorithm to give us unweighted samples.

\paragraph{Algorithm 2, Likelihood-Amplified Prior Sampling (LAPS)}

\begin{enumerate}
\item Draw $n$ samples $\theta_1, \theta_2, \ldots \theta_n$ iidly from $\PriorMeasure$.
\item For each $\theta_i$, calculate the likelihood $f(\theta_i)$.
\item Fix a constant $c > 0$, and make $\left\lceil cf(\theta_i) \right\rceil$ copies of
$\theta_i$.
\item Put all the copies of all the $\theta_i$ together in a multi-set or bag $\tilde{\theta}$.
\item Use the empirical distribution of $\tilde{\theta}$,
$\ReplicatedApproxPost$ as the approximation to $\PosteriorMeasure$.
\end{enumerate}

$\ReplicatedApproxPost$ can be viewed as forcing the weights on take on
discrete values which are separated by a scale that depends on $c$.
(Specifically, the weights must be multiples of $1/\sum_{i}{\lceil
  cf(\theta_i)\rceil}$.)  One thus has that $\ReplicatedApproxPost(A) =
\WeightedApproxPost(A) + O(1/c)$.  Making $c$ very large will however also tend
to make $\tilde{\theta}$ large and computationally awkward.  One fix for that
would be to draw a sub-sample from $\tilde{\theta}$, but if we were going to do
that, the intermediate phase of making lots of copies is unnecessary.

\paragraph{Algorithm 3, Selective Likelihood-Amplified Prior Sampling (SLIPS)}

\begin{enumerate}
\item Draw $n$ samples $\theta_1, \theta_2, \ldots \theta_n$ iidly from $\PriorMeasure$.
\item For each $\theta_i$, calculate the likelihood $f(\theta_i)$.
\item  Sample $m$ times from the $\theta_i$, with replacement, and with sampling weights proportional to $f(\theta_i)$.  Call this sample $\theta^{\dagger}$.
\item  Return the empirical distribution of $\theta^{\dagger}$,
$\SampledApproxPost$, as the approximation to $\PosteriorMeasure$.
\end{enumerate}

$\SampledApproxPost$ will approach $\WeightedApproxPost$ as $m \rightarrow
\infty$ \citep[Theorem 3]{Douc-Moulines-weighted-samples}, but $m$ can
be tuned to trade off numerical accuracy against memory demands.

\section{Implementation}

All three algorithms can be implemented in one short piece of R.

\begin{knitrout}\small
\definecolor{shadecolor}{rgb}{1, 1, 1}\color{fgcolor}\begin{kframe}
\begin{verbatim}
rposterior <- function(n, rprior, likelihood, c = 100, m = n, mode = "lips") {
    theta <- rprior(n)
    likelihoods <- apply(theta, 1, likelihood)
    if (mode == "lips") {
        integrated.likelihood <- sum(likelihoods)
        posterior.weights <- likelihoods/integrated.likelihood
    } else if (mode == "laps") {
        copies <- ceiling(c * likelihoods)
        theta <- theta[rep(1:n, times = copies), ]
        posterior.weights = rep(1, times = sum(copies))
    } else if (mode == "slips") {
        theta <- theta[sample(1:n, size = m, replace = TRUE, prob = likelihoods),
            ]
        posterior.weights = rep(1, times = m)
    }
    return(list(theta = theta, posterior.weights = posterior.weights))
}
\end{verbatim}
\end{kframe}
\end{knitrout}

Here the argument \verb_rprior_ is a function which, given an input
\verb_n_, returns a 2D array of \verb_n_ IID parameter vectors drawn from the
prior distribution, each sample on its own row.  Similarly, the
\verb_likelihood_ function must, given a parameter vector, return the
likelihood.  This code could be improved in many ways --- e.g., it makes no
attempt to take advantage of parallelized back-ends if they are available ---
but shows off the simplicity of the idea, and will suffice for an example.

\paragraph{Example: Gaussian Likelihood and Gaussian Prior} Let us take a case
where the posterior can be computed in closed form, for purposes of comparison.
The observations will have a Gaussian distribution with unknown mean $\theta$
and variance 1, and the prior measure on the measure is itself $\PriorMeasure =
N(0,1)$.  The posterior distribution is then $\PosteriorMeasure = N(x/2,1/2)$.
Figure \ref{fig:normal-normal-example-1} shows how well the scheme works in
this simple setting.

\begin{figure}
\begin{knitrout}\small
\definecolor{shadecolor}{rgb}{1, 1, 1}\color{fgcolor}
\includegraphics[width=\maxwidth]{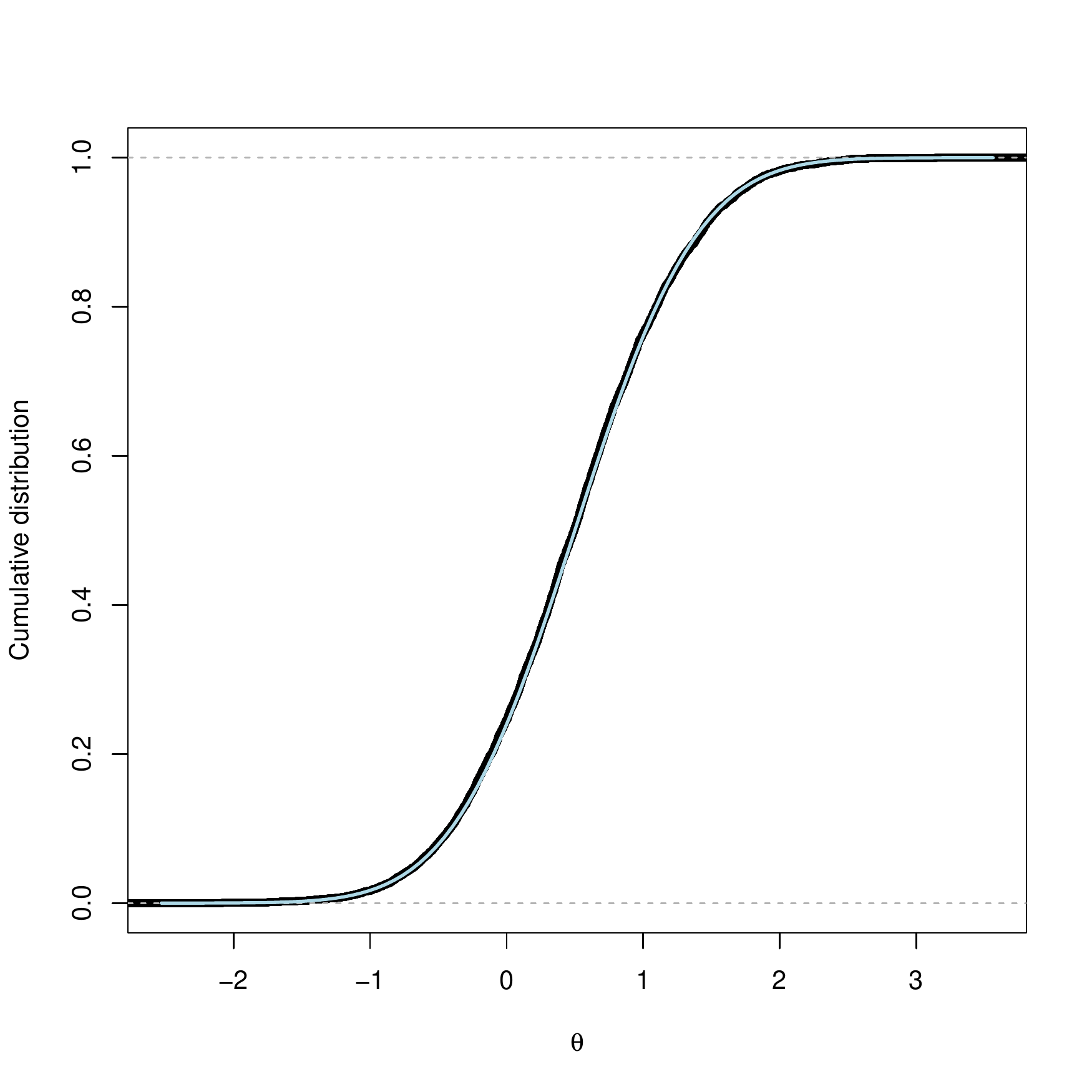} 
\end{knitrout}
\begin{knitrout}\small
\definecolor{shadecolor}{rgb}{1, 1, 1}\color{fgcolor}\begin{kframe}
\begin{verbatim}
rprior <- function(n) {
    matrix(rnorm(n, mean = 0, sd = 1), nrow = n)
}
likelihood <- function(theta) {
    dnorm(x = 1, mean = theta, sd = 1)
}
theta <- rposterior(10000, rprior = rprior, likelihood = likelihood, mode = "slips")
plot(ecdf(theta$theta), main = "", xlab = expression(theta), ylab = "Cumulative distribution",
    lwd = 5)
curve(pnorm(x, 0.5, 1/sqrt(2)), col = "lightblue", add = TRUE, lwd = 2)
\end{verbatim}
\end{kframe}
\end{knitrout}
  \caption{Example of breeding MC samples for a simple normal-normal example.
    Black line: CDF of the MC samples after breeding.  Blue line: analytic
    posterior.  Here we set $x=1$ for the observation.}
  \label{fig:normal-normal-example-1}
\end{figure}

In the preceding paragraph, the prior distribution was not that different from
the posterior.  We thus consider the case where we have taken ${10}^4$ IID
samples, so that $\overline{X}|\theta \sim N(\theta,{10}^{-4})$.  The analytic
posterior is then $\theta|\overline{X} \sim
N(\frac{{10}^{4}}{{10}^{4}+1}\overline{X},1/10001)$.  The quality of the
approximation is still quite good (Figure \ref{fig:normal-normal-example-2}),
and, of course, will only get better as $n$ increases.

\begin{figure}
\begin{knitrout}\small
\definecolor{shadecolor}{rgb}{1, 1, 1}\color{fgcolor}
\includegraphics[width=\maxwidth]{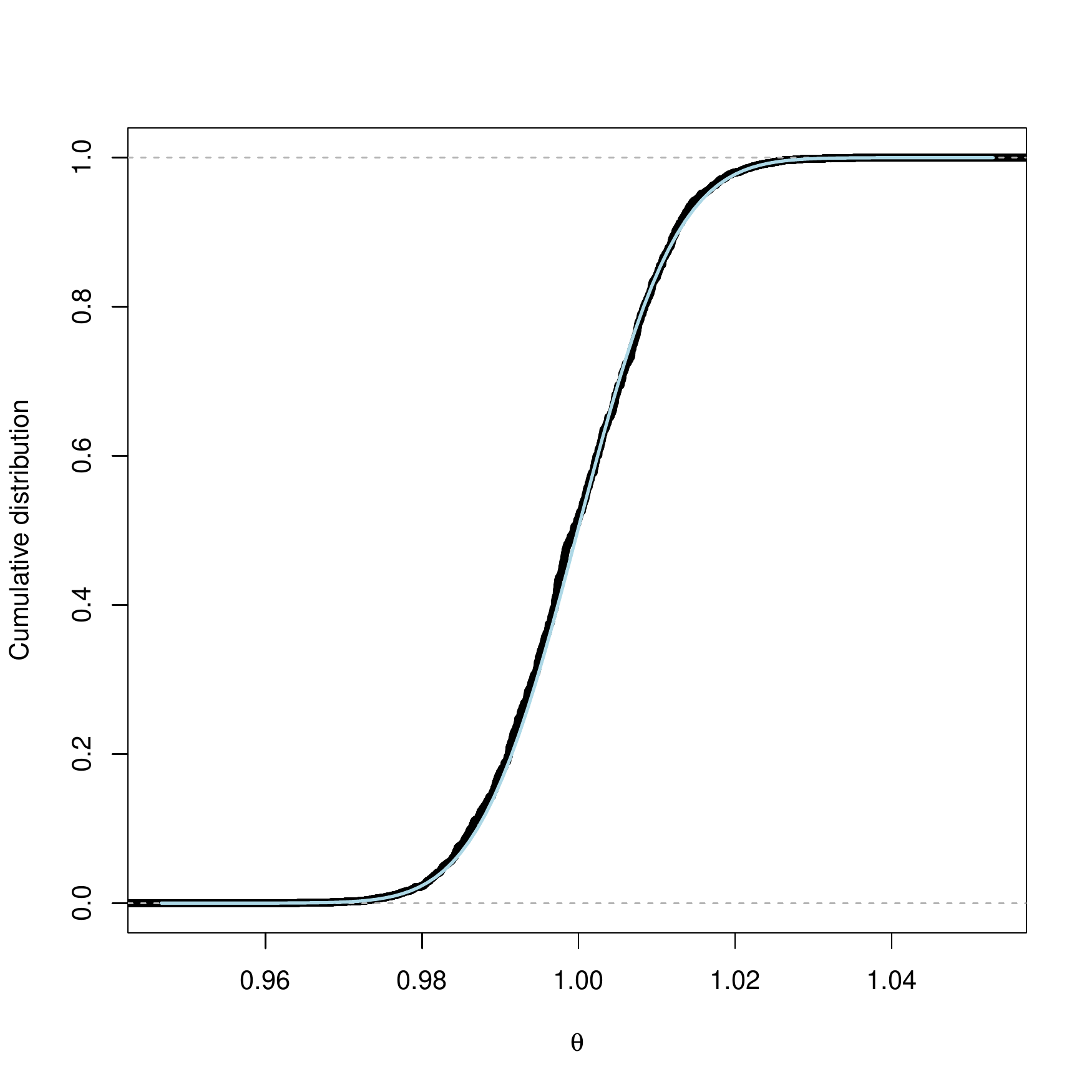} 
\end{knitrout}
\begin{knitrout}\small
\definecolor{shadecolor}{rgb}{1, 1, 1}\color{fgcolor}\begin{kframe}
\begin{verbatim}
likelihood2 <- function(theta) {
    dnorm(x = 1, mean = theta, sd = 1/sqrt(10000))
}
theta2 <- rposterior(1e+05, rprior, likelihood2, mode = "slips")
plot(ecdf(theta2$theta), main = "", xlab = expression(theta), ylab = "Cumulative distribution",
    lwd = 5)
curve(pnorm(x, (10000/(10000 + 1) * 1), sqrt(1/(10000 + 1))), col = "lightblue",
    add = TRUE, lwd = 2)
\end{verbatim}
\end{kframe}
\end{knitrout}
  \caption{Continuation of the Gaussian-Gaussian example, where we suppose that
    we have taken ${10}^4$ uncorrelated observations, each with $\sigma^2=1$,
    and $\overline{x}=1$. }
  \label{fig:normal-normal-example-2}
\end{figure}

\clearpage

\section{Discussion}

\subsection{Summary}

For any measurable $A \subseteq \Theta$, $\WeightedApproxPost(A) =
\PosteriorMeasure(A) + O_P(1/\sqrt{n})$.  Thus Algorithm 1 provides a
rapidly-converging way of estimating posterior probabilities.  The error of
approximation is controlled by $D_2(\PosteriorMeasure\|\PriorMeasure)$, or more
concretely by the ratio $\PriorMeasure(f^2)/\PriorMeasure^2(f) = 1 +
\Var{f}/\PriorMeasure^2(f)$.  This is to say, the more (prior) variance there
is in the likelihood, the worse this will work.  (The dimensionality of
$\Theta$ is not, however, directly relevant.)  Algorithms 2 and 3 simply
introduce additional approximation error, which however can be made as small as
desired either by letting $c \rightarrow \infty$ or $m \rightarrow \infty$.

\subsection{Related Work}
\label{sec:related}

These algorithms are all closely related to particle filters
\citep{Sequential-Monte-Carlo-book}, which however usually aim at approximating
a sequence of probability measures which themselves evolve randomly, and where
convergence results often rely on the ergodicity of the underlying process.
Here, however, we have only one (possibly high-dimensional) observation, and
only one posterior distribution to match.

Artificial intelligence employs many population-based optimization techniques
to explore fitness landscapes, by copying more successful solutions in
proportion to their fitness, most notably genetic algorithms
\citep{Holland-adaptation-1st, MM-on-GAs}.  All of these, however, have a vital
role for operations which introduce new members of the population,
corresponding to mutation, cross-over, sex, etc., which are necessarily lacking
in Bayesian updating\footnote{From a
strictly Bayesian point of view, innovation is merely an invitation to be Dutch
Booked, and so ``irrational'' and/or ``incoherent'' \citep{Gelman-Shalizi-philosophy}.}, and avoided in the algorithms set forth above.

Beyond these general points of inspiration, \citet{Rubin-on-SIR}
proposed what is basically Algorithm 3 as a way of doing multiple
imputation, as opposed to sampling from the posterior of the
parameters.
\citet{Cappe-et-al-population-monte-carlo} is an even more precise
anticipation, though their use of sequential importance sampling means
they aren't just sampling from the prior, with attendant loss of
simplicity and analytical tractability.

\subsection{Advantages}
\label{sec:pros}

Best practices in Markov chain Monte Carlo counsel discarding all the samples
from an initial burn-in period, running diagnostic tests to check converge to
the equilibrium, etc.  (Admittedly, some doubt the value of these steps: see
\citealt{Geyer-on-MCMC}.)  None of this is needed or even applicable here: all
that's needed is for $n$ to be large enough that $\WeightedApproxPost$ and/or
$\ReplicatedApproxPost$ is a good approximation to the posterior distribution
$\PosteriorMeasure$.  Moreover, there is no need to pick or adjust proposal
distributions.  Indeed, there aren't really any control settings that need to
be {\em tuned}, since it's straightforward that $n$, $c$ and $m$ should all be
as large as possible.

Implementing this scheme requires only the ability to draw IID samples
from the prior $\PriorMeasure$, and to compute the likelihood function
$f$.  The former tends to be straightforward, at least when the prior
is expressed hierarchically.  Actually, examination of the proofs will
confirm that we do not even need to calculate $f$; it would be enough
to calculate something proportional to the likelihood, $r(x)
f(\theta)$, where the factor of proportionality $r(x) > 0$ and is
strictly constant in $\theta$.  That being said, calculating the
likelihood can be difficult for complex models, but this is outside
the scope here\footnote{If the difficulty in calculating $f(\theta;x)$
  comes from integrating out latent variables, say $Y \in \Upsilon$,
  so that $f(\theta;x) = \int_{\Upsilon}{p(x|y;\theta) p(y|\theta)
    dy}$, there is a straightforward work-around if $p(x|y;\theta)$ is
  computationally tractable and $p(y|\theta)$ is easy to simulate
  from.  Formally expand the parameter space to $\Theta^{\prime} =
  \Theta \times \Upsilon$, drawing $\theta_i$ from $\PriorMeasure$ and
  $Y_i$ from $p(y|\theta_i)$.  Then use $p(x|y;\theta)$ as the
  likelihood, and apply the posterior approximation only to sets of
  the form $A \times \Upsilon$.}.

One special advantage of this scheme is that it lends itself readily to
parallelization, which MCMC does not.  Drawing iidly from the prior is
``embarrassingly'' parallel.  Calculating $f(\theta_i)$ is also embarrassingly
parallel, requiring no communication between the different $\theta_i$, as is
making copies of them.  Every step in algorithm 2 is embarrassingly parallel,
though with large $c$ we might need a lot of memory in each node.  Algorithms 1
and 3 require us to compute $\sum_{i=1}^{n}{f(\theta_i)}$, which does require
communication, but finding the sum of $n$ numbers is among the first
parallelized algorithms taught to beginners \citep[\S
2.1]{Burch-intro-to-parallel-and-distributed}, so this might not be much of a
drawback.

\subsection{Disadvantages}
\label{sec:cons}

The great advantage of all three algorithms is that the only distribution they
ever need to draw from is the prior, and priors are typically constructed to
make that easy.  The corresponding disadvantage is that they can spend a lot of
time evaluating the likelihood of parameter values which have in fact very low
likelihood, and so end up making a small contribution to the posterior.  MCMC,
by contrast, preferentially spends most of its time in regions of high
posterior density.

Three observations may, however, lessen this contrast.  First, while
the {\em output} of MCMC follows the posterior distribution, its {\em
  proposals} do not, and the likelihood must be evaluated once for
each proposal.  Thus it is not clear that, in general, MCMC has a
computational advantage on this score.  Second, while I have not
explored it in this brief note, it should be possible to short-cut the
evaluation of the likelihood of very hopeless parameter points, as in
the ``go with the winners'' of
\citet{Grassberger-go-with-the-winners}\footnote{Grassberger traces
  the idea of randomly eliminating bad parameter values, and making
  more copies of the good ones, to \citet{Kahn-on-Monte-Carlo}.  The
  latter, in turn, gave this trick the arresting name of ``Russian
  Roulette and splitting'', and attributed it to an unpublished paper
  by John von Neumann and Stanislaw Ulam.}, or the ``austerity'' of
\citet{Korattikara-et-al-austerity}.  Third, the problematic case is
where most of the posterior probability mass is concentrated in a
region of very low prior mass.  Numerical evidence from trying out
more complicated situations than the Gaussian prior-and-likelihood
suggests that this can be a serious issue, which is why I do not
advocate these algorithms, in their present form, as a replacement for
MCMC (note that the non-Markov chain ``population Monte Carlo'' of
\citet{Cappe-et-al-population-monte-carlo} does {\em not} suffer from
this drawback).  It is, however, at least arguable that what goes
wrong in these situation isn't that these algorithms converge slowly,
but that the priors are bad, and should be replaced by less misleading
ones \citep{Gelman-Shalizi-philosophy}.

\subsubsection*{Acknowledgments}

Thanks to my students in 36-350, introduction to statistical computing, for
driving me to this way of explaining Bayes's rule, and to John Miller for many
valuable conversations about the connections between adaptive population
processes and Monte Carlo.  I was supported while writing this by grants from
the NSF (DMS1207759, DMS1418124) and from the Institute for New Economic
Thinking.  \S\S \ref{sec:asymptotics} and \ref{sec:laplace} were worked out
during breaks in the proceedings while serving as a juror in January 2014, and
so also supported by jury duty pay.

\bibliographystyle{crs}
\bibliography{locusts}

\end{document}